\documentclass[12pt]{iopart}
\usepackage{iopams}  

\usepackage{graphicx}

\def\rs{r_\mathrm{s}}

\def\Zel{{Z_e}}

\def\beq{\begin{equation}}
\def\eeq{\end{equation}}

\def\reff#1{(\ref{#1})}

\def\subsc#1{{\mathrm{#1}}}

\def\N3d{N_\subsc{3D}}

\def\vekt#1{{\mathbf{#1}}}

\def\vektr{\vekt{r}}

\def\vektE{\vekt{E}}

\def\Vxcsigma{V_{\mathrm{xc}\sigma}}
\def\Vxcisigma{V_{\mathrm{xc}i\sigma}}

\def\imagi{\mathrm{i}}
\def\diff{\mathrm{d}}

\def\VH{V_\mathrm{H}}
\def\VHisigma{V_{\mathrm{H}i\sigma}}

\def\rs{r_\mathrm{s}}

\begin{document}

\title[Modeling core-hole screening in jellium clusters using density functional theory]{Modeling core-hole screening in jellium clusters using density functional theory}

\author{D Bauer}

\address{Institut f\"ur Physik, Universit\"at Rostock, 18051 Rostock, Germany}
\ead{dieter.bauer@uni-rostock.de}
\begin{abstract}
The screening of a 2p core-hole in Na clusters is investigated using density functional theory applied to an extended jellium model with an all-electron atom in the center. The study is related to recent experiments at the free electron laser at DESY in which photoelectron spectra from mass-selected, core-shell ionized metal clusters have been recorded. Relaxed and unrelaxed binding energies as well as Kohn-Sham orbital energies are calculated in Perdew-Zunger self-interaction-corrected exchange-only local spin-density approximation for valence and 2p core electrons in Na clusters up to 58 atoms.  The relaxed binding energies follow  approximately the metal-sphere behavior. The same behavior is seen in the experiment for sufficiently big clusters, indicating perfect screening and that the relaxation energy due to screening goes to the photoelectron. Instead, calculating the kinetic energy of the photoelectrons using unrelaxed binding energies or Kohn-Sham orbital energies yields wrong results for core-shell electrons. 
The screening {\em dynamics} are investigated using time-dependent density functional theory. It is  shown that screening occurs on two time scales, a core-shell-dependent inner-atomic and an inter-atomic valence electron time scale. In the case of Na 2p ionization the remaining electrons in the 2p shell screen within tens of attoseconds while the screening due to cluster valence electrons occurs within several hundreds of attoseconds. The screening time-scales may be compared to the photon energy and cluster size-dependent escape times of the photoelectron in order to estimate whether the photoelectron is capable of picking up the relaxation energy or whether the residual system is left in an excited state.  

\end{abstract}

\maketitle

\section{Introduction}
Core-holes are created when matter is irradiated with energetic photons.  The net binding energy is obtained from the kinetic energy of the photoelectron upon subtracting the photon energy. As the target size increases from the isolated atom via molecules and clusters to, ultimately, bulk matter,  the shift of the net binding energy contains information about the screening capability of the
other electrons in the system and allows to investigate, e.g., metal-to-nonmetal transitions. In fact, x-ray photoelectron spectroscopy (XPS) is well-established and work has been extensively devoted to measure core-level binding energy shifts in solids and at surfaces \cite{egel}. However, the systematic study of core-level shifts as a function of the target size became possible only recently because of the necessity of having  size-selected clusters  and tunable, powerful x-ray sources such as free-electron lasers available. 

From the theoretical point of view, the calculation of core-level binding energies in many-electron systems is non-trivial both conceptually and computationally \cite{egel,johans}. The study of atoms, clusters or bulk, in which suddenly a core-electron is removed by single-photon ionization, has a long history, starting probably with Slater's transition state theory \cite{hadji,olovsson} and various kinds of self-consistent field methods ($\Delta$SCFs) \cite{puska,jones}. $\Delta$SCF-approaches are based on energy balances between electronic configurations, and it is not always clear whether relaxed or unrelaxed configurations should be taken. Here, ``relaxed'' means that the other electrons are allowed to assume the new, energetically most favorable configuration---albeit without {\em filling} the core-hole. ``Unrelaxed'' means that the other electrons are kept frozen. It is intuitively clear that the unrelaxed energy difference between final and initial configuration should be considered if the photoelectron escapes too rapidly to notice the relaxation dynamics of the other electrons \cite{manne,lundqvist,privalov}. 

Methodologically, self-consistent field methods like Hartree-Fock (HF) or density functional theory (DFT) (see, e.g., \cite{dftbook}) are usually employed for the calculation of energy differences between final and initial configurations in $\Delta$SCF approaches. As both methods introduce orbitals and orbital energies, the interesting question arises how these orbital energies are related to the (relaxed or unrelaxed) binding energies \cite{perdewzunger,jones,mundt}. Clearly, the energy difference depends on initial {\em and} final state while the orbital energies do not, unless fractional occupation numbers, as in  Slater's transition state theory, are introduced.

We use an extended Na-cluster jellium model with an {\em all-electron} Na atom in the center in order to have core-levels in the system at all. This model may be viewed as a hybrid of the pure cluster jellium model \cite{ekardt} and bulk-models in which atoms are immersed in homogeneous electron gas \cite{puska,puska2}.  We employ DFT to calculate relaxed and unrelaxed energy differences between the initial ground state and the final configuration with either a valence electron or a 2p-core electron removed. The results as a function of the cluster size  (up to 58 atoms) are compared with the prediction of the metal sphere-model. 
Time-dependent DFT (TDDFT) (see \cite{tddftbook} for a state-of-the-art account) is used to investigate the screening {\em dynamics} after the sudden removal of a 2p electron. We apply our model in this first study to Na clusters (instead of Pb$^-$-clusters in the experiments \cite{senz,njpbahn}) in order to keep the number of inner electrons in  the embedded atom and thus the numerical effort managable.

The paper is organized as follows. The basic methodology and theoretical background is introduced in section \ref{basictheory}. The DFT results on the size-dependence of binding energies and the TDDFT results on the screening dynamics are presented in   section \ref{results}. We conclude in section \ref{concl}.

Atomic units $\hbar=m_e=|e|=4\pi\epsilon_0=1$ are used, unless noted otherwise.

\section{Basic theory  and models}\label{basictheory}
We use DFT in exchange-only local spin-density approximation (xLSD) with Perdew-Zunger (PZ) self-interaction correction (SIC) \cite{perdewzunger}. The electronic Kohn-Sham (KS) equation reads
\beq \epsilon_{i\sigma}\varphi_{i\sigma} = \biggl( T + V(r) + \VH + \Vxcsigma - {\{ \VHisigma + \Vxcisigma \}} \biggr)\varphi_{i\sigma}.   \label{KS1}\eeq
Here, $\epsilon_{i\sigma}$ is the energy of KS orbital $\varphi_{i\sigma}$, $i=1,2,\ldots N_\sigma$ with $N_\sigma$ the number of electrons with spin-projection $\sigma=\uparrow,\downarrow$. $T$ is the kinetic energy operator, $V(r)$ is the spherically symmetrical external potential introduced below. $\VH$ and $\Vxcsigma$ are the Hartree potential and the exchange-correlation (xc) potential in xLSD approximation, respectively (see, e.g., \cite{dftbook}). Both are functionals of the (spin) density. The term in the curly bracket  is the PZ-SIC, i.e., in the KS equation for orbital $\varphi_{i\sigma}$ the Hartree-xc potential  $\VHisigma + \Vxcisigma$ evaluated with the corresponding spin density $n_{i\sigma}=|\varphi_{i\sigma}|^2$ is subtracted from the full Hartree-xc potential $\VH + \Vxcsigma$ evaluated with
\beq n=\sum_\sigma n_\sigma = n_\uparrow + n_\downarrow, \qquad n_\sigma=\sum_{i=1}^{N_\sigma}  n_{i\sigma}. \eeq
It is known that the PZ-SIC applied to xLSD significantly improves the results as compared to uncorrected xLSD \cite{perdewzunger,vydrov}.  
The central-field approximation is applied to  $\VH + \Vxcsigma - {\{ \VHisigma + \Vxcisigma \}}$ in \reff{KS1} so that the single-particle orbital angular momentum quantum numbers $\ell$ and magnetic quantum numbers $m$ remain good quantum numbers.  Although the PZ-SIC in general improves the uncorrected KS results quantitatively, there have been several critical issues discussed in the literature \cite{koerz,vydrov}, one of them being  the orbital-dependent effective Hamiltonian in \reff{KS1}, leading to non-orthogonal KS orbitals. In our calculations we therefore force the KS orbitals to be orthogonal.  An extended version of the Qprop code \cite{qprop,qpropwww} is used to solve \reff{KS1} and---for the dynamics in section \ref{tddftresults}---its time-dependent version (in which $\epsilon_{i\sigma}$ is replaced by $\imagi \partial_t$ and the respective adiabatically time-dependent potentials are used \cite{tddftbook}). 

\subsection{Cluster jellium model with central all-electron atom}\label{clustermodel}
The external potential $V(r)$ in \reff{KS1} in this work is 
\beq V(r)= - \frac{Z}{r}  + \left\{ \begin{array}{ccc} \displaystyle -\frac{A-1}{2 R} \left( 3-\frac{r^2}{R^2} \right) & \mathrm{for} & r<R \\
           \displaystyle   -\frac{A-1}{r}  & \mathrm{for} & r\geq R \end{array} \right. . \label{potential} \eeq
The first term $-Z/r$ is the pure Coulomb potential of the central atom for which all electrons are taken into account. This atom provides the ionizable core-shell. The other $A-1$ atoms in the cluster are treated in jellium approximation.  The jellium potential is that of a homogeneously charged sphere of radius 
\beq R=A^{1/3}\rs,\label{radius}\eeq 
mimicking the net ionic background of the cluster, as ``seen'' by the cluster valence electrons. A Wigner-Seitz radius for Na of $\rs=4$ was used. For $A=1$ the potential \reff{potential} clearly reduces to the single-atom case. 

\begin{figure}\begin{center}
\includegraphics[width=0.5\textwidth]{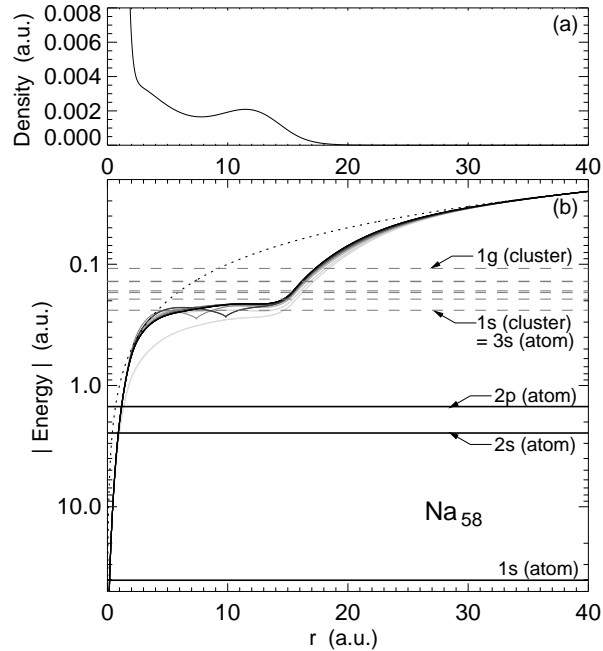}
\end{center}
\caption{Extended jellium model for a Na$_{58}$ cluster. (a) Total density (strong increase in center because of explicitly treated all-electron Na atom there). (b) KS potentials for the various PZ-KS orbitals (gray to black), $1/r$ dotted; horizontal lines indicate KS levels (central atomic ones solid, cluster levels dashed). \label{Na58resdensandpot}}
\end{figure}

Figure~\ref{Na58resdensandpot} shows the results for electron density, the PZ-KS potentials, and the KS levels for the neutral closed-shell Na$_{58}$ cluster. Because all electrons in the central atom are considered there is a  peak in the electron density in the center. What would be the 3s level in the isolated Na atom is now the 1s jellium cluster level. The higher the jellium level the closer the respective KS eigenvalue is to the  one in the corresponding pure valence electron jellium model. Note that the PZ-SIC yields the correct $1/r$-behavior of the KS potentials, unlike pure xLSD.

\subsection{Determination of the core-shell binding energies}
The experimental observable is the kinetic energy $\epsilon_\mathrm{kin}$ of photoelectrons which are emitted upon the irradiation of clusters with x-rays of photon energy $\hbar\omega$. Energy conservation requires
\beq  E_0 + \hbar \omega = E_\mathrm{f} + \epsilon_\mathrm{kin}, \qquad E_0 < E_\mathrm{f} < 0 \eeq
so that the binding energy can be determined as
\beq \Delta E =  E_\mathrm{f} - E_0 = \hbar \omega - \epsilon_\mathrm{kin}, \qquad  \Delta E > 0. \label{experimentalbinding} \eeq
$E_0<0$ is the initial energy of the target. In what follows we assume that the target is initially in its ground state. However,  it is far from obvious what $E_\mathrm{f}$ is. As long as Auger decay occurs on time-scales longer than the photoelectron escape times the final target configuration of interest to us is a cluster in which one atom has a core-hole.\footnote{Auger decay will only later lead to secondary, low-energy electrons which, in fact, are seen in experiments \cite{egel}.} We assume that the photon energy is such that most likely one of the 2p electrons is removed. If the escape time is longer than the screening time the photoelectron may pick up the relaxation energy
\beq \Sigma=   E_\mathrm{unrlxd}- E_\mathrm{rlxd} > 0, \label{relaxenerg} \eeq
and the net binding energy is
\beq \Delta E_\mathrm{rlxd} =  E_\mathrm{rlxd} - E_0 = E_\mathrm{unrlxd} - E_0 - \Sigma . \label{bindingenergy1}\eeq
On the other hand, if the photoelectron escapes before the system is relaxed, the binding energy reads
\beq \Delta E_\mathrm{unrlxd}= E_\mathrm{unrlxd} - E_0
 =  E_\mathrm{rlxd} - E_0 +  \Sigma > \Delta E_\mathrm{rlxd}. \eeq

According to Koopmans' theorem the unrelaxed energy difference equals the Hartree-Fock orbital energy, $-\epsilon^\mathrm{HF}=E_\mathrm{unrlxd}^\mathrm{HF} - E_0^\mathrm{HF}>0$. The fact that in the ultra-high photon energy-limit the centroid of the photoelectron spectrum is located at $\epsilon_\mathrm{kin}=\epsilon^\mathrm{HF}+\hbar\omega$ is known as the Manne-{\AA}berg theorem \cite{manne} or the Lundqvist sum rule \cite{lundqvist}. However, Koopmans' theorem does not hold in DFT. Instead, the ``non-Koopmans energy'' tends to cancel a part of the relaxation energy \reff{relaxenerg} \cite{perdewzunger} so that KS orbital energies $|\epsilon|=-\epsilon$ are shifted towards $\Delta E_\mathrm{rlxd}$.  Whether $-\epsilon$ is above or below $\Delta E_\mathrm{rlxd}$ depends on the xc energy functional used. Introducing fractional occupations $0\leq f\leq 1$, Janak's theorem establishes a connection between the {\em relaxed} binding energy and the {\em relaxed} KS orbital energies $\epsilon(f)$ (see, e.g., \cite{perdewzunger}).

Using the extended cluster jellium model introduced in section \ref{clustermodel} we determined the binding energy 
\beq \Delta E_{i\sigma} = E_{i\sigma} - E_0 \eeq
 of a core-shell electron $i\sigma$ by switching the fractional occupation number of the respective KS orbital $f_{i\sigma}$ from 1 to 0. Without relaxing the final configuration with the core-hole we obtain
\beq  \Delta E_{{i\sigma},\mathrm{unrlxd}} = E_{{i\sigma},\mathrm{unrlxd}} - E_0, \eeq 
and with relaxation
\beq  \Delta E_{{i\sigma},\mathrm{rlxd}} = E_{{i\sigma},\mathrm{rlxd}} - E_0 <  \Delta E_{{i\sigma},\mathrm{unrlxd}}. \eeq 
The relaxation energy 
\beq \Sigma_{{i\sigma}} = E_{{i\sigma},\mathrm{unrlxd}} - E_{{i\sigma},\mathrm{rlxd}}= \Delta E_{{i\sigma},\mathrm{unrlxd}} - \Delta E_{{i\sigma},\mathrm{rlxd}}, \eeq
in general, depends on the orbital from which the electron is removed. Clearly, in the case of the removal of a valence electron the relaxation energy is expected to be very small because mainly the electrons {\em above} the level from which the electron is removed contribute to screening. On the other hand, for the screening by the cluster valence electrons only the net positive, almost point-like hole of  charge $|e|$ counts, and it should not matter much from {\em which} inner-atomic shell a core-electron is removed. Possible differences in the relaxation energy are expected to be due to other core-shell electrons. We call this {\em inner-atomic screening}, in contrast to {\em inter-atomic screening} by cluster valence electrons.
 
A widely used approach in the DFT community is to use the KS orbital energies as a zeroth-order approximation for the binding energy of electrons. However, the initial state KS orbital energy may be above or below $\Delta E_{{i\sigma},\mathrm{rlxd}}$, depending on the xc potential used \cite{perdewzunger}.   Energy differences $\Delta E$ are less sensitive and, even more so, binding energy {\em shifts}, i.e., the difference of energy differences. 

\subsection{Metal sphere model}
The energy required to remove one electron from a metal sphere of radius $R$ which, initially, has $\Zel$ negative elementary {\em excess} charges $-|e|$ reads
\beq \Delta E_{\mathrm{metsph},\Zel}  = E_{\Zel-1} - E_\Zel = W + \frac{1-2\Zel}{2R}. \label{metsph}\eeq
Here, $W$ is the work function. In the bulk limit $R\to\infty$ one has $\Delta E_{\mathrm{metsph}}=W$, independent of $\Zel$.
Equation~\reff{metsph} can be simply derived by calculating the initial and final energies. As the interior of the metal sphere is, by definition, field-free one has
\beq
E_\Zel = \frac{1}{8\pi} \int_R^\infty\diff r \ 4\pi r^2 \vektE^2(r)= \frac{1}{8\pi} \int_R^\infty\diff r \ 4\pi r^2 \left( \frac{\Zel\vektr}{r^3} \right)^2 = \frac{\Zel^2}{2R} \eeq
(with $\vektE(r)$ the electric field) so that
\beq 
E_{\Zel-1} - E_\Zel = \frac{(\Zel-1)^2-\Zel^2}{2R} = \frac{1-2\Zel}{2R}. \eeq
The work $W$ needed to remove an electron from the bulk is added ``by hand,'' leading to \reff{metsph}. Improved metal sphere models yield also $W$ and helped to resolve the ``image charge paradox'' \cite{seidl}\footnote{Calculating the work needed to move the electron from the metal sphere surface to infinity using the image charge-method yields a formula different from \reff{metsph}. Only if the work  is taken into account that is required to move the electron from inside the metal sphere through the surface, \reff{metsph} is recovered.}.

\begin{figure}\begin{center}
\includegraphics[width=0.5\textwidth]{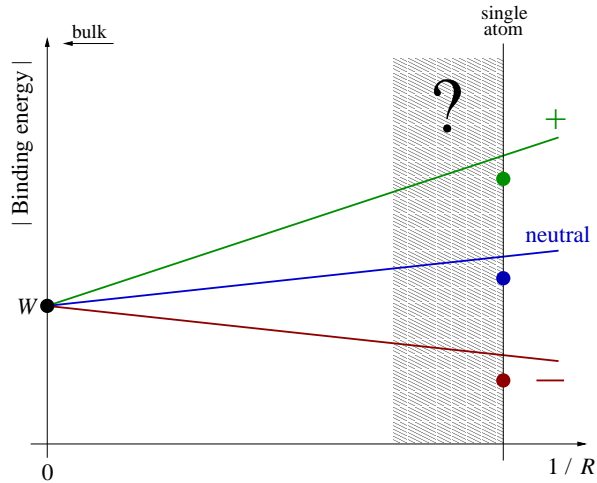}
\end{center}
\caption{Qualitative sketch of the metal sphere model result eq.~\reff{metsph} for the binding energy of cluster anions, neutral clusters, and cluster cations vs the inverse cluster radius. Starting at $1/R\to 0$ (bulk) with the work function $W$ the slope is negative for electron removal from anions (dark red, $-$), positive for neutrals (blue), and steeper positive for cations (green, $+$). Possible single, isolated atom or ion results are indicated by filled circles, which, in general do not lie on the metal sphere lines. The transition from the perfect metal behavior to the single, isolated atom or ion result occurs in the hatched region.  \label{workf}}
\end{figure}

Figure \ref{workf} summarizes schematically the prediction of the metal sphere model. Starting from the bulk value $W$ for the binding energy for $1/R\to 0 $ eq.~\reff{metsph} predicts a linear behavior in $1/R$ with a negative slope for initially negatively charged clusters, a positive slope for initially neutral clusters, and increasingly steep positive slopes for more and more positively charged clusters. While the cluster radius is---apart from a small spill-out---well-defined for big clusters this is not the case for small clusters.  In the following we will use  \reff{radius} down to the single, isolated atom (where $R=\rs$) and hence plot the respective binding energy vs $1/\rs$. Even with the uncertainty in the definition of the single atom or ion radius kept in mind, the binding energy for the single atom or ion in general does not lie on the metal sphere line predicted by \reff{metsph}. Hence, there is a transition region (hatched area in Fig.~\ref{workf}) in which the binding energy is expected to deviate from the metal sphere line towards the single atom or ion result (filled colored circles in Fig.~\ref{workf}).

\section{Results and discussion}  \label{results}
Table \ref{tab1} summarizes the PZ-SIC xLSD results for the smallest and largest system we consider in this work, namely the single, isolated Na atom ($A=1$), and the Na$_{58}$ cluster ($A=58$). One sees that in the single atom-case the removal of the 3s$\uparrow$ valence electron and subsequent relaxation  gives $\Delta E_{\mathrm{3s}\uparrow,\mathrm{rlxd}}=0.191$, in good agreement with the experimental ionization potential $0.189$\footnote{A pure xLSD calculation yields only $0.179$, showing that the PZ-SIC greatly improves the result.}.
The same procedure for a 2p$\downarrow$ yields a binding energy $\Delta E_{\mathrm{2p}\downarrow,\mathrm{rlxd}}=1.395$, i.e., $\simeq 38$\,eV, in very good agreement with the experimental result for the $^3$P$_2$ XPS peak \cite{pered}\footnote{Because spin-orbit coupling is neglected, we do not differentiate fine-structure in our KS calculations.}. The removal of a  2p$\uparrow$-electron gives the slightly different and higher binding energy $\Delta E_{\mathrm{2p}\uparrow,\mathrm{rlxd}}=1.403$.
Looking at the KS orbital energies one recognizes that $-\epsilon_{\mathrm{3s}\uparrow,\mathrm{rlxd}}=0.186$ is slightly below $\Delta E_{\mathrm{3s}\uparrow,\mathrm{rlxd}}=0.191$. Hence, Koopmans' theorem is almost fulfilled for the valence electron, which is not surprising because relaxation effects are expected to be small if a loosely bound outer electron is removed. 

The difference between KS orbital energy and relaxed energy difference is larger for the 2p-electron removal: $-\epsilon_{\mathrm{2p}\downarrow}=1.452>\Delta E_{\mathrm{2p}\downarrow,\mathrm{rlxd}}=1.395$. Obviously, the relaxation energy does {\em not} fully cancel the ``non-Koopmans energy'' \cite{perdewzunger}.

For Na$_{58}$ we see in the right Table \ref{tab1} that the binding energy of the valence electron  $\Delta E_{\mathrm{1g},\mathrm{rlxd}}$ is reduced compared to the atomic case, as expected from the metal sphere model for an initially neutral cluster. The binding energy $\Delta E_{\mathrm{2p},\mathrm{rlxd}}=1.308$ also shifts to lower values with increasing cluster size.  However, the absolute value of the KS orbital energy increased with respect to the atomic value. Hence, naively estimating the photoelectron peak position in an XPS spectrum to be $\epsilon_\mathrm{kin}= \epsilon_{\mathrm{2p}} + \hbar\omega $ is not only inaccurate but even may predict peak shifts into the wrong direction.

\begin{table}
\caption{\label{tab1}PZ-SIC xLSD results for the single, isolated Na atom (left) and the Na$_{58}$ cluster (right). The subscripts $a$ and $c$ on the right hand side indicate whether the level belongs to the central atom or jellium cluster, respectively. The Na$_{58}$ cluster is a spin-neutral system with closed shells  (i.e., $A=58$ is a ``magic number'') so that binding energies are independent on whether a spin-up or spin-down electron is removed.  }

\begin{flushright}
\parbox{0.4\textwidth}{\begin{tabular}{rrr}\br
Na$_{1}$ & orbital    & energy  \\ \mr
&1s$\uparrow$ & $-40.333$ \\
&2s$\uparrow$ & $-2.425$ \\
&2p$\uparrow$ & $-1.453$ \\
&3s$\uparrow$ & $-0.186$ \\ \mr
&1s$\downarrow$ & $-40.333$ \\
&2s$\downarrow$ & $-2.424$ \\
&2p$\downarrow$ & $-1.452$ \\ \mr
&$E_0$ &   $-162.217$ \\
&$E_{\mathrm{2p}\downarrow,\mathrm{rlxd}}$ & $-160.822$\\
&$E_{\mathrm{3s}\uparrow,\mathrm{rlxd}}$ & $-162.027$ \\
&$\Delta E_{\mathrm{2p}\downarrow,\mathrm{rlxd}}$ & $1.395$ \\
&$\Delta E_{\mathrm{3s}\uparrow,\mathrm{rlxd}}$ & $0.191$ \\  \br
&  &  \\
&  &  \\
&  &  
\end{tabular}} 
\parbox{0.4\textwidth}{\begin{tabular}{rrr}\br
 Na$_{58}$ &orbital    & energy  \\ \mr
 &1s$_{a}$ & $-40.367$ \\
 &2s$_{a}$ & $-2.465$ \\
 &2p$_{a}$& $-1.493$ \\ \mr
 &3s$_{a}$ / 1s$_{c}$ & $-0.240$ \\ 
 &1p$_{c}$& $-0.194$ \\ 
 &1d$_{c}$ & $-0.166$ \\ 
 &2s$_{c}$ & $-0.171$ \\ 
 &2p$_{c}$ & $-0.139$ \\ 
 &1f$_{c}$ & $-0.138$ \\ 
 &1g$_{c}$ & $-0.108$ \\ \mr
 &$E_0$ &  $-351.215$ \\
 &$E_{\mathrm{2p},\mathrm{rlxd}}$ & $-349.906$\\
 &$E_{\mathrm{1g},\mathrm{rlxd}}$ & $-351.101$ \\
 &$\Delta E_{\mathrm{2p},\mathrm{rlxd}}$ & $1.308$ \\
 &$\Delta E_{\mathrm{1g},\mathrm{rlxd}}$ & $0.114$ \\ \br
\end{tabular}}
\end{flushright}
\end{table}

\subsection{Energies vs $1/R$}
\begin{figure}\begin{center}
\includegraphics[width=0.6\textwidth]{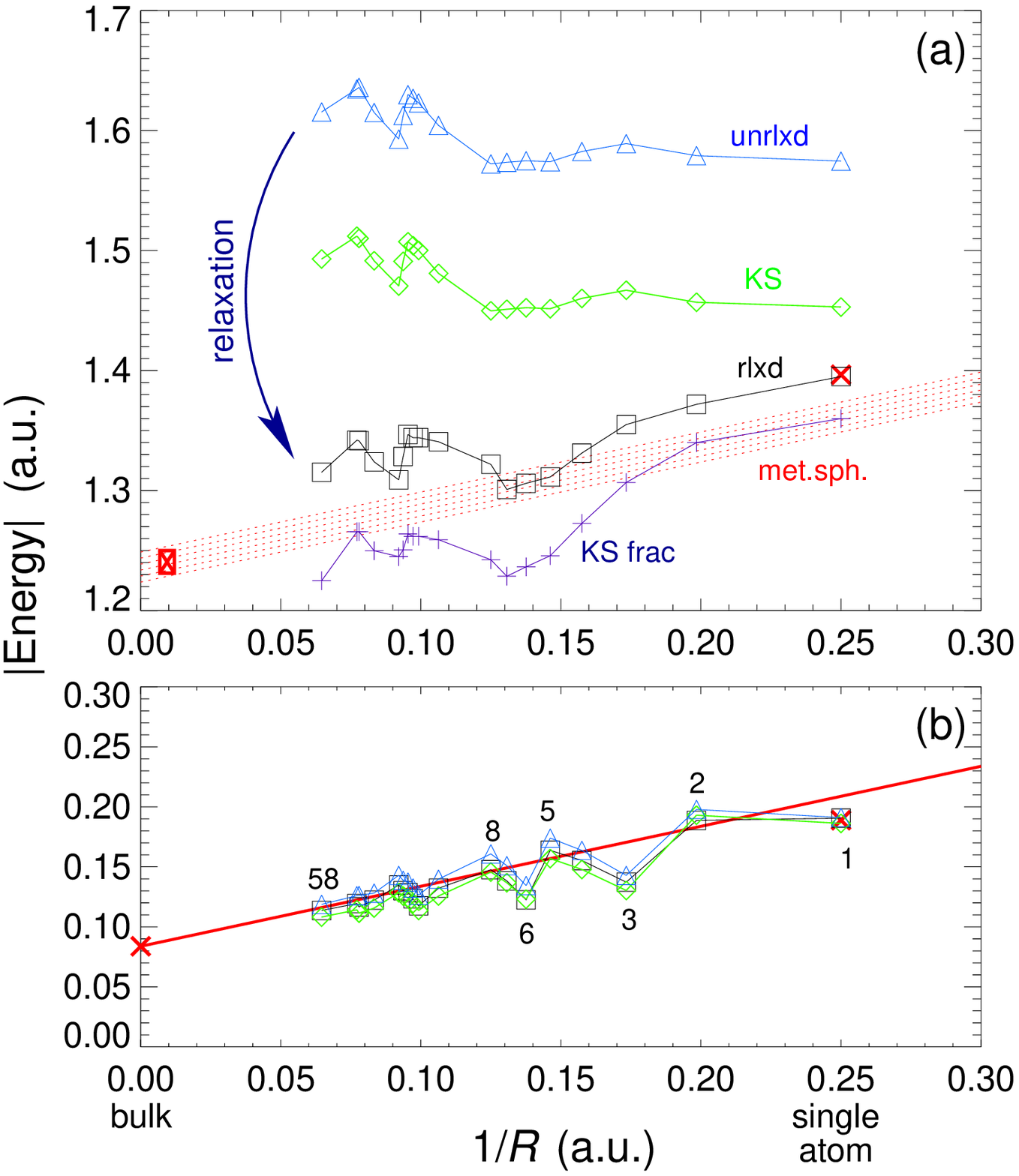}
\end{center}\caption{Relaxed energy-differences (black $\Box$), unrelaxed energy-differences (blue $\bigtriangleup$), and absolute value of KS orbital energies (green $\Diamond$) vs inverse cluster radius $R^{-1}=(A^{1/3}\rs)^{-1}$ for the removal of an atomic 2p electron (a) and a cluster valence electron (b). The small annotations close to the symbols in (b) indicate the number of atoms $A$ in the cluster.  Red crosses indicate experimental data points, red lines [several dashed in (a) and one solid in (b)] the metal sphere-model prediction \reff{metsph} for $\Zel=0$. For the valence case (b) there is almost no difference between relaxed binding energy, unrelaxed binding energy, and the KS orbital energy. However,  for the removal of a 2p core-shell electron the differences are large and increase with increasing cluster size. The difference between unrelaxed and relaxed binding energy is the relaxation energy, as indicated by the curved arrow in (a). The purple + symbols in (a) show the minimum absolute value of the 2p KS orbital energy for the relaxed configuration as the 2p ground state fractional occupation number is varied between 1 and 0 (see text for discussion).   \label{sizedep1}}
\end{figure}

Figure~\ref{sizedep1}b shows $\Delta E_{\mathrm{valence},\mathrm{rlxd}}$,
$\Delta E_{\mathrm{valence},\mathrm{unrlxd}}$, and $|\epsilon_{\mathrm{valence}}|$ vs $1/R$. We observe the well-known zig-zag behavior of the binding energy because of shell-closures \cite{ekardt,engel,seidlmbbrack} before the metal-sphere result is approached for increasing cluster size.  For the removal of a valence electron the KS orbital energies $|\epsilon_{\mathrm{valence}}|$ are very close to $\Delta E_{\mathrm{valence},\mathrm{rlxd}}$, and both develop towards the metal sphere behavior as $1/R$ decreases. The relaxation energy $\Sigma_{\mathrm{valence}}= \Delta E_{\mathrm{valence},\mathrm{unrlxd}} - \Delta E_{\mathrm{valence},\mathrm{rlxd}}$ is very small.

The typical valence electron result for binding energies in clusters shown in Fig.~\ref{sizedep1}b is known for decades \cite{ekardt,engel,seidlmbbrack}. The qualitatively new aspect coming into play when core-electrons are removed is the reorientation of the $AZ+\Zel-1$ electrons after the emission of the core-shell photoelectron. There are (at least) three options:
\begin{itemize}
\item[(i)] the photoelectron is slow enough to gain the relaxation energy which is set free due to the screening of the core-hole by the other electrons but fast enough to escape before the core-hole itself decays;
\item[(ii)] the photoelectron escapes so rapidly that it cannot pick up the relaxation energy;
\item[(iii)]  the photoelectron is slow enough to gain energy not only from relaxation but also from Auger-like core-hole decay.
\end{itemize}
Signatures in XPS spectra that would support option (iii) were not observed in experiments \cite{senz,njpbahn}. Instead, slow ``secondary'' electrons are measured, which are likely generated via Auger decay of the core-hole {\em after} the photoelectron is already gone. Indeed the estimated photoelectron escape times are sub-femtosecond for the cluster sizes and photon energies considered in this work and thus faster than Auger decay \cite{rander}. 

Option (i) is supported by the experiment because the experimental binding energies \reff{experimentalbinding} for sufficiently big clusters follow again the metal sphere prediction with only the work function shifted to the respective bulk binding energy.

Option (ii) would manifest itself as an {\em increased} binding energy of the photoelectron as compared to the situation where all the relaxation energy goes to the photoelectron. Instead, the experimentally measured binding energies in \cite{senz,njpbahn} drop {\em below} the metal-sphere result, which ultimately must be so because the single ion value (Pb$^-$ in \cite{senz,njpbahn}) lies below the metal-sphere line. Hence, one has to distinguish (at least) two effects here: the ability of the other electrons to screen and the ability of the photoelectron to pick up the relaxation energy due to screening. The first one seems to be relevant in the experiments \cite{senz,njpbahn} so that indeed the metal-to-nonmetal transition in Pb$_A^-$ around $A=20$ could be probed. The situation for initially neutral  Na clusters is quantitatively different.  In fact, we find---apart from the shell-oscillations---little deviation from the metal-sphere behavior down to the single-atom limit in Fig.~\ref{sizedep1}b, in agreement with experiment \cite{bowlan}.

Figure~\ref{sizedep1}a shows the PZ-SIC xLSD result for the removal of a 2p core-shell electron from the central, all-electron Na atom. Again, the relaxed energy difference follows approximately  the trend predicted by the metal sphere model, although with substantial fluctuations on the eV-level ($1$\,eV $\simeq 0.037$\,a.u.). These fluctuations are also correlated with the filling of electronic shells but in a less obvious way as compared to the valence case where maxima in the (absolute value of the) binding energy occur at half-filled and full shells. It would be desirable to move towards bigger cluster in order to see whether our model really approaches the metal-sphere result for the screened 2p core-hole. However, it is numerically very demanding to proceed significantly towards the bulk limit. Note that $1/R=0.05$ already amounts to $A=125$. 

Figure~\ref{sizedep1}a clearly shows the importance of the relaxation energy $\Sigma_{\mathrm{2p}}=\Delta E_{\mathrm{2p},\mathrm{unrlxd}} - \Delta E_{\mathrm{2p},\mathrm{rlxd}}$ (i.e., the difference between blue triangles and black squares). As expected, the relaxation energy increases with the cluster size. For the biggest cluster treated in this work, Na$_{58}$, we have $\Sigma_{\mathrm{2p}}^{(58)}=0.3$, i.e., $8.2$\,eV. 

The KS 2p-orbital energies $|\epsilon_\mathrm{2p}|$ of the ground state (green diamonds) are {\em between} the relaxed and the unrelaxed energies, as it is expected from PZ-SIC xLSD \cite{perdewzunger}. We checked how the KS orbital energy shifts as the fractional occupation number $f_\mathrm{2p}$ is decreased from 1 to 0 in the always {\em relaxed} system. We observed that $|\epsilon_\mathrm{2p}(f_\mathrm{2p})|$ decreases below $\Delta E_{\mathrm{2p},\mathrm{rlxd}}$ (purple + symbols in Fig.~\ref{sizedep1}a) and then, for  $0.2 <f_\mathrm{2p} \leq 0$ proceeds towards the value for the core-shell-ionized system. If the unknown exact xc potential could be employed the orbital energy would {\em jump} discontinuously as $f_\mathrm{2p}$ changes from $0<\varepsilon \ll 1$ to $0$. This is the  so-called ``derivative discontinuity'' \cite{perdewdiffdisc}. The PZ-SIC xLSD mimics this behavior but smoothes-out the discontinuity.   We see that the metal sphere result in  Fig.~\ref{sizedep1}a lies between the relaxed binding energy and the minimum  KS 2p-orbital energies $\min|\epsilon_\mathrm{2p}(f_\mathrm{2p})|$. Fractional occupation numbers are related to Slater's transition state theory where the idea is to consider a (relaxed) configuration  ``half way'' between initial and final state. Indeed, $\epsilon_\mathrm{2p}(f_\mathrm{2p}\simeq 0.5) + \hbar\omega$ would predict a photoelectron peak in XPS spectra close to what is expected from the metal-sphere model whereas $\epsilon_\mathrm{2p}(f_\mathrm{2p}=1) + \hbar\omega$  would be completely off and---apart from fluctuations---erroneously {\em increasing} with increasing cluster size.

Concerning the metal-sphere result for the 2p core-level binding energy, it should be noted that the 2p bulk limit is not precisely known. Moreover, the experimental peaks are typically splitted over $\simeq 0.4$\,eV because of fine-structure. We estimated the 2p bulk value  from extrapolating the binding energy obtained from the large-cluster XPS spectrum in \cite{pered}. As a consequence, the core-level  metal sphere result in  Fig.~\ref{sizedep1}a   may have an uncertainty of $\pm 1$\,eV $\simeq \pm 0.04$\,atomic units (red dashed area).

The experimental results for the core-level binding energies as a function of the cluster size \cite{senz,njpbahn} follow the metal-sphere prediction down to a certain minimum cluster size and then deviate, ultimately approaching the single atom or ion result, which in the experiment with Pb$^-$ clusters was {\em below} the metal sphere binding energy for the 5d and 4f core-shells (unlike for the 2p shell in Na above). A too rapid escape of the photoelectron would lead to an increased binding energy. It thus seems that the relaxation energy that is set free due to screening goes indeed to the photoelectron and that the deviation from the metal sphere-model is an element-specific electronic structure-effect. In fact, it is expected that from a certain minimal cluster size on the spherical jellium approximation breaks down. However, our extended jellium model has the exact single atom-limit, so it may deviate at a too small cluster size from the metal sphere line but it ultimately will. The only exception arises when the single atom or ion binding energy itself lies---perhaps by chance---on the metal sphere line. Indeed, for sodium this is almost the case, as is seen in Fig.~\ref{sizedep1}.

\subsection{TDDFT study of the screening dynamics}  \label{tddftresults}
\begin{figure}\begin{center}
\includegraphics[width=0.7\textwidth]{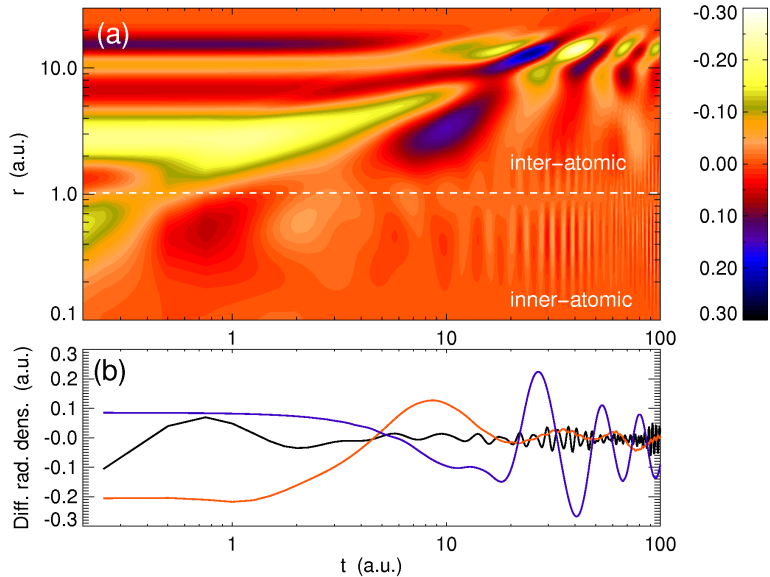}
\end{center}
\caption{(a) Contour plot of the difference in the radial spin density $4\pi r^2[n^\uparrow(r,t) - n^\uparrow_{\mathrm{2p}\downarrow,\mathrm{rlxd}}(r)]$ in Na$_{58}$ vs time $t$ and radius $r$ (both logarithmically) after the sudden removal of a 2p$\downarrow$ electron at time $t=0$. (b) Line-outs for $r=0.51$ (black), $r=2.51$ (orange), $r=14.01$ (blue), showing the different time-scales involved, i.e., ``inner-atomic'' and ``inter-atomic'', respectively, as indicated in (a).  \label{diffdens}}
\end{figure}

In order to understand why the relaxation energy that is set free due to screening goes to the photoelectron we determined the relevant time scales. To that end we removed at time $t=0$ instantaneously a  2p$\downarrow$ electron from the central, all-electron Na atom in the Na$_{58}$-cluster, and propagated the KS orbitals thereafter. Figure~\ref{diffdens} shows the difference in the radial spin densities \beq 4\pi r^2[n^\uparrow(r,t) - n^\uparrow_{\mathrm{2p}\downarrow,\mathrm{rlxd}}(r)]\eeq
as a function of time $t$ and radial position $r$. The spin-down density $n^\downarrow$ shows a very similar behavior and is therefore not shown. At small $r<1$ mainly the charge density of the inner electrons (up to the 2p-shell from which the photoelectron has been removed) screens on a very fast time-scale ($\simeq 2$\,a.u.\ $\simeq 48$\,as) and continues to oscillate. This is simple to understand: after the sudden creation of the core-hole, the KS orbitals experience a new potential in which they are not eigenstates anymore. If the potential was stationary one would expand the KS orbitals in the new eigenstates, and the oscillation periods would be given by the inverse eigenenergy-separations. In TDDFT the situation is more involved because the potential is nonlinear, i.e., it depends on the KS orbital densities. Moreover, we cannot employ linear response theory here because the change in the fractional occupation $f_{\mathrm{2p}\downarrow}$ number is not small but unity. Nevertheless, the energy width spanned by the inner-electronic levels is such that it yields the screening time-scale visible in  Fig.~\ref{diffdens}.  As can be inferred from the density oscillations for bigger $r$, the valence electrons screen on a time-scale of $\simeq 20$\,a.u.\ $\simeq 0.5$\,fs, i.e.\ still sub-fs, in agreement with Ref.~\cite{borisov}.\footnote{In Ref.~\cite{borisov}, the dynamics after the sudden immersion of a {\em negative} charge in a jellium cluster was investigated.} The decay of the density oscillations seen in  Fig.~\ref{diffdens} is probably underestimated because spontaneous emission, electron ion collisions, energy transfer to ionic degrees of freedom, and perhaps other possible dissipation channels are absent in our TDDFT approach. 

The time-dependent calculations clearly reveal that the screening dynamics, and thus the relaxation, are fast enough to provide the escaping photoelectron with the relaxation energy. For instance, the kinetic energy of a 2p electron in a Na$_{58}$-cluster that is removed by a $40$\,eV photon is  $\simeq 4.6$\,eV, and it takes $\simeq 0.6$\,fs to escape from the cluster. Hence, it certainly has enough time to pick up the relaxation energy from its own shell (and the shells below), and even the time-scale for the valence electron screening is comparable.

\section{Conclusions} \label{concl}
A jellium model with a central all-electron atom has been introduced in order to model recent experiments on core-shell ionization of metal clusters  as a function of the cluster size. Density functional theory in exchange-only local spin-density approximation with a Perdew-Zunger self-interaction correction has been employed to calculate the self-consistent electron configurations of the model applied to sodium clusters. The {\em relaxed} binding energy follows the trend predicted by the metal sphere model and confirmed experimentally. However, shell-structure-related fluctuations more pronounced than those for valence electron-removal are observed. Both the {\em unrelaxed} binding energy and the initial, ground state Kohn-Sham orbital energy are completely off, demonstrating a substantial relaxation energy, the importance of screening, and the irrelevance of initial-state core-shell Kohn-Sham eigenvalues. Predicting a photoelectron peak by  simply adding the photon energy to the initial Kohn-Sham core-level energy is, in general, doomed to fail. Instead, for the exchange-correlation potential used in this work Kohn-Sham eigenvalues evaluated for fractional occupation numbers (in the spirit of Slater's transition state theory) are closer to the relaxed binding energy and metal-sphere result. 

Time-dependent density functional theory has been applied in order to investigate the screening {\em dynamics} after the sudden removal of a Na 2p  core-electron. {\em Inner-atomic} screening due to electrons in the same and other core-shells of Na occurs within a few tens of attoseconds. {\em Inter-atomic} screening by the  cluster valence electrons occurs rather on the hundreds-of-attoseconds time scale. The inner-atomic time scale is, of course, core-shell dependent and may be even faster for deeper-lying shells and heavier elements while the screening time scale  in metal clusters due to valence electrons is less sensitive and depends mostly on the number of valence electrons per atom. Whether the photoelectron ``has enough time'' to pick up the relaxation energy depends on the ratio of escape time to relaxation time. The higher the photon energy and the smaller the cluster the higher should be the net binding energy. However, the deviation from the metal-sphere model observed so far experimentally is most likely dominated by electronic-structure effects. In  order to verify this assertion future studies will go beyond the jellium model.

\section*{Acknowledgments}
Fruitful discussions with K-H Meiwes-Broer are gratefully acknowledged. This work has been supported by the Collaborative Research Center SFB 652 at the University of Rostock.


\section*{References}
\begin{thebibliography}{10}
\bibitem{egel} Egelhoff W F  1987 Core-level binding energy shifts at surfaces and in solids {\em Surface Science Reports} {\bf 6} 253 


\bibitem{johans} Johansson B and M{\aa}rtensson N 1980 Core-level binding energy shifts for the metallic elements {\em Phys.\ Rev.\ B} {\bf 21} 4427 

\bibitem{hadji} Hadjisavvas N and Theophilou A 1985 Rigorous formulation of Slater's transition-state theory for excited states {\em Phys.\ Rev.\ A} {\bf 32}  720

\bibitem{olovsson} Olovsson W, G\"oransson C, Marten T and Abrikosov  I A 2006 Core-level shifts in complex metallic systems from first principle {\em Phys.\ Stat.\ Sol.\ B} {\bf 243} 2447  

\bibitem{puska} Puska M J and Nieminen R M 1982 Density-functional calculations of Auger and x-ray photoemission shifts for metallic elements {\em Physica Scripta} {\bf 25} 708 

\bibitem{jones} Jones R O and  Gunnarsson O 1989  The density functional formalism, its applications and prospects {\em Rev.\ Mod.\ Phys.} {\bf 61}  689


\bibitem{manne} Manne R and {\AA}berg T 1970 Koopmans' theorem for inner-shell ionization {\em Chem.\ Phys.\ Lett.} {\bf 7} 282

\bibitem{lundqvist} Lundqvist B I 1969 Characteristic structure in core electron spectra of metals due to the electron-plasmon coupling {\em Phys.\ kond.\ Mat.} {\bf 9} 236

\bibitem{privalov} Privalov T, Gel'mukhanov F and {\AA}gren H 2001 Role of relaxation and time-dependent formation of x-ray spectra  {\em Phys.\ Rev.\ B} {\bf 64} 165115

\bibitem{dftbook} Parr R G and Yang W 1989 {\em Density-Functional Theory of Atoms and Molecules (International Series of Monographs on Chemistry 16)} (Oxford: Oxford University Press)

\bibitem{perdewzunger} Perdew J P and Zunger A 1981 Self-interaction correction to density-functional approximations for many-electron systems {\em Phys.\ Rev.\ B} {\bf 23} 5048 %

\bibitem{mundt} Mundt M, K\"ummel S, Huber B and Moseler M 2006 Photoelectron spectra of sodium clusters: the problem of interpreting Kohn-Sham eigenvalues 2006 {\em Phys.\ Rev.\ B} {\bf 73} 205407


\bibitem{ekardt} Ekardt W 1984 Work function of small metal particles: self-consistent spherical jellium-background model {\em Phys.\ Rev.\ B} {\bf 29} 1558


\bibitem{puska2} Puska M J, Nieminen R M , Manninen M 1981 Atoms imbedded in an electron gas: immersion energies {\em Phys.\ Rev.\ B} {\bf 24} 3037 


\bibitem{tddftbook} Marques M A L, Maitra N T, Nogueira F M S, Gross E K U and  Rubio A (eds) 2012  {\em Fundamentals of Time-Dependent Density Functional Theory (Lecture Notes in Physics 837)} (Berlin Heidelberg: Springer)

\bibitem{senz} Senz V {\em et al.} 2009 Core-hole screening as a probe for a metal-to-nonmetal transition in lead clusters {\em Phys.\ Rev.\ Lett.} {\bf 102} 138303 %

\bibitem{njpbahn} Bahn J {\em et al.} {\em this NJP Focus Issue}


\bibitem{koerz} K\"orzd\"orfer T, K\"ummel S and Mundt M 2008 Self-interaction correction and the optimized effective potential {\em J.\ Chem.\ Phys.} {\bf 129} 014110

\bibitem{vydrov} Vydrov O A and Scuseria G E 2005 Ionization potentials and electron affinities in the Perdew-Zunger self-interaction corrected density-functional theory {\em  J.\ Chem.\ Phys.} {\bf 122} 184107

\bibitem{qprop} Bauer D and Koval P 2006 Qprop: A Schr\"odinger-solver for intense laser-atom interaction {\em Comput.\ Phys.\ Comm.} {\bf 174} 396

\bibitem{qpropwww} http://www.qprop.de

\bibitem{seidl} Seidl M and Perdew J P 1994 Size-dependent ionization energy of a metallic cluster: Resolution of the classical image-potential paradox {\em Phys.\ Rev.\ B} {\bf 50} 5744 %

\bibitem{pered} Peredkov S {\em et al.} 2007 Free nanoscale sodium clusters studied by core-level photoelectron spectroscopy {\em Phys.\ Rev.\ B} {\bf 75} 235407 %

\bibitem{engel} Engel E and Perdew J P 1991 Theory of metallic clusters: asymptotic size dependence of electronic properties {\em Phys.\ Rev.\ B} {\bf 43} 1331

\bibitem{seidlmbbrack} Seidl M, Meiwes-Broer K-H and Brack M 1991 Finite-size effects in ionization potentials and electron-affinities of metal clusters  {\em  J.\ Chem.\ Phys.} {\bf 95} 1295



\bibitem{rander} Rander T, Schulz J, Huttula M,  M\"akinen A, Tchaplyguine M, Svensson S, \"Ohrwall G, Bj\"orneholm O, Aksela S  and Aksela H  2007 Core-level electron spectroscopy on the sodium dimer Na 2p level {\em Phys.\ Rev.\ A} {\bf 75}  032510


\bibitem{bowlan} Bowlan J, Liang A and de Heer W A 2011 How metallic are small sodium clusters?  {\em Phys.\ Rev.\ Lett.} {\bf 106} 043401 %

\bibitem{perdewdiffdisc} Perdew J P, Parr R G Levy M and Balduz J L 1982 Density-functional theory for fractional particle number: derivative discontinuity of the energy {\em Phys.\ Rev.\ Lett.} {\bf 49} 1691

\bibitem{borisov} Borisov A, S\'anchez-Portal D, D\'iez Mui\~no R, Echenique P M 2004 Building up the screening below the femtosecond scale {\em Chem.\ Phys.\ Lett.} {\bf 387} 95











\end{thebibliography}
\end{document}